\documentclass[aps,prb,reprint,twocolumn,superscriptaddress,showpacs]{revtex4}
\usepackage{graphicx}
\usepackage{hyperref}
\usepackage{color}

\begin{document}
\title{A hybrid quantum circuit consisting of a superconducting flux qubit coupled to both a spin ensemble and a transmission-line resonator}

\author{Ze-Liang Xiang}
%\email{09110190024@fudan.edu.cn}
\affiliation{Department of Physics and State Key Laboratory of Surface Physics, Fudan University, Shanghai 200433, China}
\affiliation{Advanced Science Institute, RIKEN, Saitama 351-0198, Japan}

\author{Xin-You L\"u}
\affiliation{Advanced Science Institute, RIKEN, Saitama 351-0198, Japan}

\author{Tie-Fu Li}
\affiliation{Institute of Microelectronics and Tsinghua National Laboratory of Information Science and Technology, Tsinghua University, Beijing 100084, China}
\affiliation{Beijing Computational Science Research Center, Beijing 100084, China}

\author{J. Q. You}
\affiliation{Department of Physics and State Key Laboratory of Surface Physics, Fudan University, Shanghai 200433, China}
\affiliation{Beijing Computational Science Research Center, Beijing 100084, China}
\affiliation{Advanced Science Institute, RIKEN, Saitama 351-0198, Japan}

\author{Franco Nori}
\affiliation{Advanced Science Institute, RIKEN, Saitama 351-0198, Japan}
\affiliation{Department of Physics, The University of Michigan, Ann Arbor, Michigan 48109, USA}

\date{\today}

\begin{abstract}
We propose an experimentally realizable hybrid quantum circuit for achieving a strong coupling between a spin ensemble and a transmission-line resonator via a superconducting flux qubit used as a data bus. The resulting coupling can be used to transfer quantum information between the spin ensemble and the resonator. In particular, in contrast to the direct coupling without a data bus, our approach requires far less spins to achieve a strong coupling between the spin ensemble and the resonator (e.g., three to four orders of magnitude less). This proposed hybrid quantum circuit could enable a long-time quantum memory when storing information in the spin ensemble, and allows the possibility to explore nonlinear effects in the ultrastrong-coupling regime.
\end{abstract}


\maketitle

\section{Introduction}\label{sec:1}

Cavity quantum electrodynamics (QED) involving the interaction between light and matter is widely utilized in implementing quantum communication and quantum information processing. It can be realized in several mixed systems, such as atom-cavity devices and spin-cavity systems, which have been studied for many years. An atomic system has stable energy levels that can be used to represent the different states of qubits.~\cite{Lukin:2003,Blatt:2008} Moreover, the coherence time of isolated atoms (or spins) is long because of their weak interaction with the environment. However, due to the small dipole moment and weak fields in the cavity, the coupling strength $g$ in these systems is usually not in the strong coupling regime corresponding to $g\gg\kappa,\gamma$, where $\kappa$ and $\gamma$ are the decay rates of the cavity and the atomic system, respectively. Remarkable progress has been made on superconducting (SC) circuits,~\cite{You:2005,Wendin:2007,Clarke:2008} where the SC qubit behaves as an artificial atom. Such SC circuits promise good scalability and allow robust control, storage and readout, owing to their strong interaction with external fields.~\cite{You:2003} SC circuits consisting of SC qubits coupled to a SC resonator, such as a transmission-line resonator, are often called circuit QED, which were widely used in quantum technologies in recent years.~\cite{Schoelkopf:2008} The strong coupling, even ultrastrong coupling,~\cite{Ashhab:2010} between the SC qubit and the resonator has also been experimentally achieved.~\cite{Chiorescu:2004,Wallraff:2004,Niemczyk:2010} However, compared with atomic systems, SC qubits have relatively short coherence times.

Recently, intense effort has been devoted to coupling atomic system with SC qubits to form hybrid quantum circuits, aiming to combine ``the best of two worlds'' (see \onlinecite{Xiang:2013} and references therein). There are two different approaches to couple these two subsystems. In one approach, both of the atomic system and the SC qubit couple to a common SC resonator, which plays the role of a data bus.~\cite{Rabl:2007,Tordrup:2008,Verdu:2009,Petrosyan:2009,Zhang:2009,Yang:2011,Kubo:2011,Amsuss:2011} Due to the weak coupling between a single atom (or spin) and the SC resonator, an ensemble with a large number $N$ of atoms (or spins) is employed for enhancing the coupling strength by a factor of $\sqrt{N}$. Recently, this approach has been experimentally demonstrated (see Ref.~\onlinecite{Kubo:2011}). However, in the presence of inhomogeneous broadening, the high density of atoms (or spins) would lead to short coherence times. ~\cite{Acosta:2009,Stanwix:2010}

In the other approach, an atomic system directly couples to a flux qubit via the magnetic field produced by the qubit.~\cite{Marcos:2010,Zhu:2011,Hoffman:2011,Hummer:2012,Stephens:2012} The coupling strength can be about three orders of magnitude stronger than using a transmission-line resonator as the data bus. However, the controllability of this approach is not good and the states in both subsystems are easily affected by each other due to their direct coupling.

In this paper, we propose a hybrid quantum circuit consisting of a SC flux qubit coupled to a spin ensemble and a transmission-line resonator. Nitrogen-vacancy (NV) centers in diamond are used as the spin ensemble in our approach because of their long coherence times, even at room temperature.~\cite{Wrachtrup:2006,Hanson:2008,Doherty:2012} Therefore such a spin ensemble can be used as a quantum memory in the hybrid quantum circuit. Note that Ref.~\onlinecite{Twamley:2010} proposed to resonantly couple a flux qubit in a transmission-line resonator with a single NV center, while now we are considering an ensemble of NV centers. Moreover, here we study how to achieve a strong effective coupling between the spin ensemble and the resonator, by adiabatically eliminating the degrees of freedom of the flux qubit. With this strong effective coupling we can transfer the quantum information from the spin ensemble to the photon states in the resonator, which can be used as flying qubits for quantum communication. In addition, in our proposed circuit, the flux qubit shares a segment with the central line of the resonator, so as to achieve a very strong coupling strength between the flux qubit and the resonator.~\cite{Bourassa:2009} The effective coupling between the spin ensemble and the resonator via this flux qubit is stronger than the direct coupling between the resonator and the same number of spins without using the flux qubit. Therefore, for a given value of the coupling strength, less spins in the ensemble are required in our approach, as compared to the direct-coupling approach. This design has the potential to achieve a larger quantum coherence time for the spin ensemble, which acts as a quantum memory. Furthermore, we also discuss the case when the coupling strength between the flux qubit and the resonator (spin ensemble) reaches the ultrastrong coupling regime. In this case, the effective coupling between the spin ensemble and the resonator is much increased, but nonlinear terms appear in the resulting effective Hamiltonian. These nonlinear terms rapidly reduce the fidelity of the quantum state transfer.

This paper is organized as follows. In Sec.~\ref{sec:2}, we describe our proposed hybrid quantum circuit and give the total Hamiltonian of the whole system. Then, we derive the effective interaction Hamiltonian between the spin ensemble and the resonator by considering the strong coupling regime in Sec.~\ref{sec:3} and the ultrastrong coupling regime in Sec.~\ref{sec:4}. Finally, a brief discussion and conclusion are given in Sec.~\ref{sec:5}.

\section{Proposed hybrid quantum circuit}\label{sec:2}

\begin{figure}[tbp]
\includegraphics[width=3.45in]{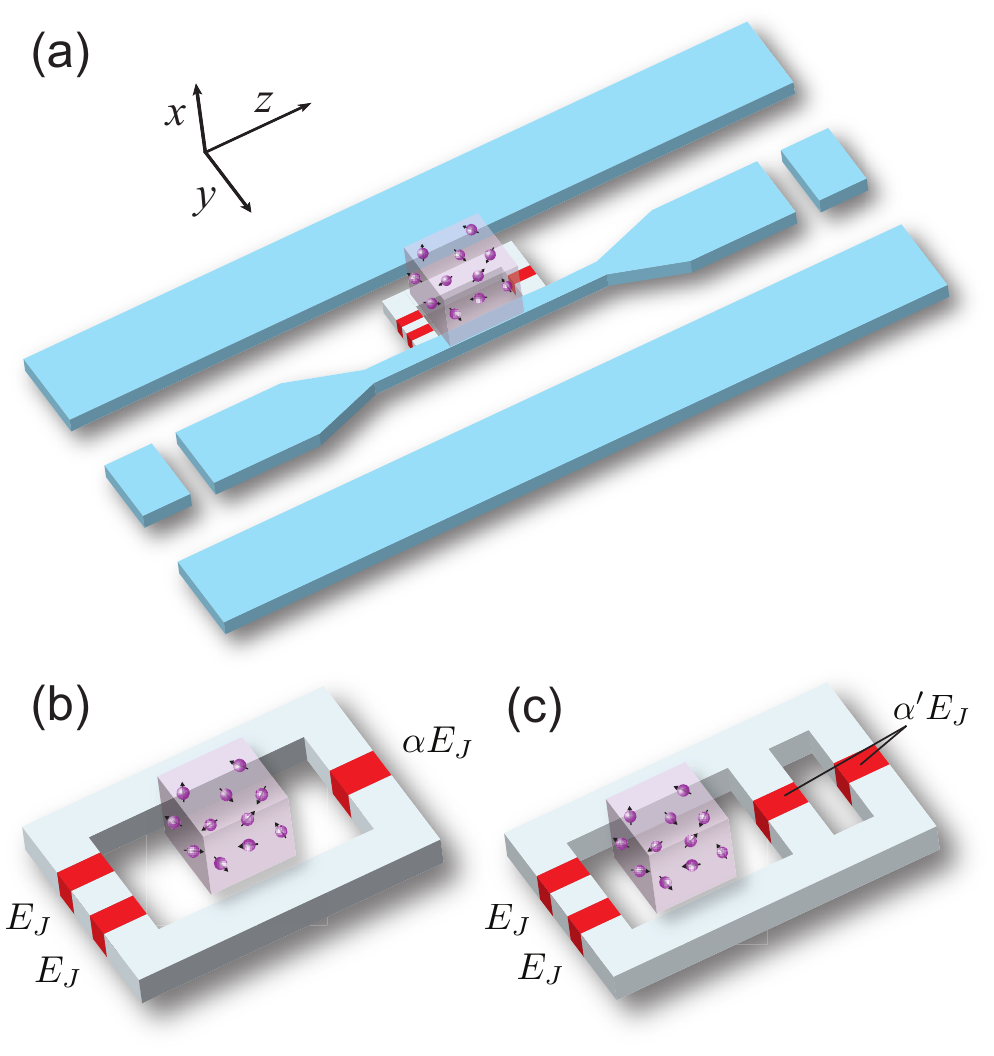}
\caption{(Color online) (a) Schematic diagram of the hybrid quantum circuit. (b) Schematic diagram of the subsystem consisting a three-junction flux qubit and a spin ensemble. (c) Schematic diagram of the subsystem consisting a tunable four-junction flux qubit and a spin ensemble. In these three diagrams, the blue part refers to the transmission-line resonator, the silver and red parts denote the superconductor and insulator parts of the flux qubit, respectively, and the light purple part shows the spin ensemble. In a three-junction flux qubit (b), two of these junctions are identical, with the same Josephson coupling energy $E_J$, while the other junction has lower Josephson coupling energy $\alpha E_J$, where $\alpha<1$. In a tunable four-junction flux qubit (c), the small junction is replaced by a SQUID, whose effective Josephson coupling energy $\alpha^{\prime}E_J$ ($\alpha^{\prime}<1$) could be adjusted by an external magnetic field threading through the SQUID loop.}
\label{fig1}
\end{figure}

We consider the hybrid quantum circuit shown in Fig.~\ref{fig1}(a), which is composed of a spin ensemble, a SC flux qubit, and a one-dimensional cavity formed by a transmission-line resonator. The spin ensemble, composed of $N$ identical and noninteracting spins, is placed inside or slightly above the qubit loop (see Fig.~\ref{fig1}). The flux qubit shares a segment with the central line of the resonator at an antinode of the standing wave of the current in the transmission-line, as shown in Fig.~\ref{fig1}(a).

We will first derive the Hamiltonian of the hybrid system consisting of a flux qubit and a spin ensemble [see Fig.~\ref{fig1}(b) or \ref{fig1}(c)], and then obtain the total Hamiltonian of the proposed hybrid quantum circuit in Fig.~\ref{fig1}(a).

%===============================================
%===============================================

\subsection{Flux qubit coupled to a spin ensemble}

\begin{figure}[h]
\includegraphics[width=3.2in]{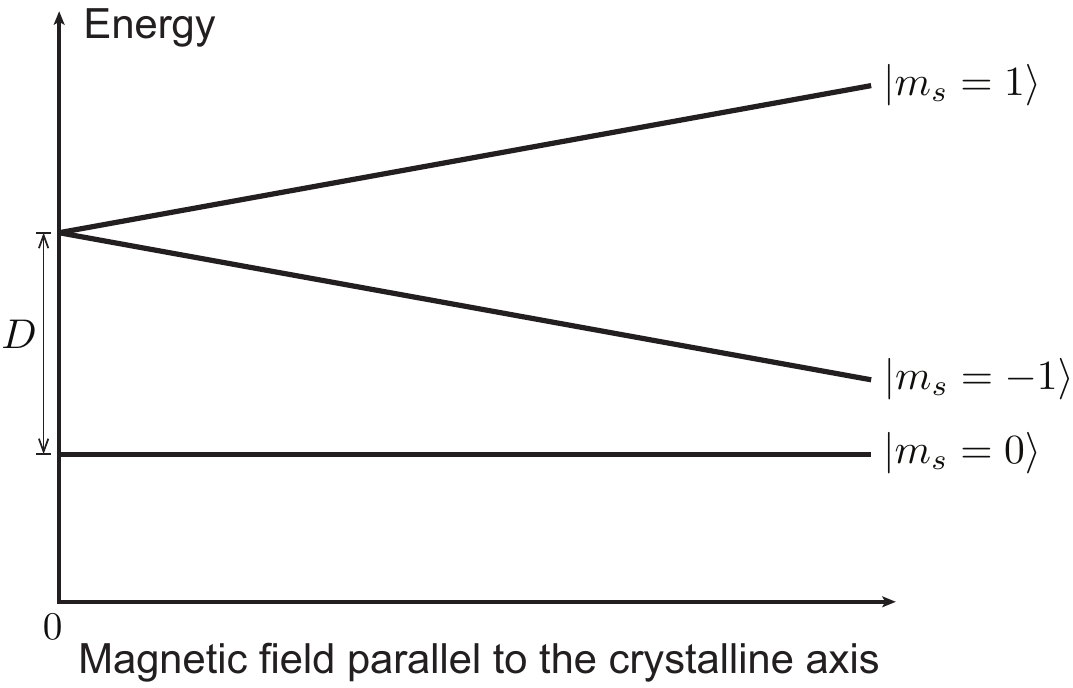}
\caption{The ground electronic-spin states of the NV center in the presence of an external magnetic field parallel to the crystalline axis.}
\label{fig2}
\end{figure}

We use NV centers as the spin ensemble, whose spin-1 triplet sublevels of the electronic ground state have a zero-field splitting $\Delta\approx 2\pi\times 2.87$ GHz between the $m_s=0$ and $m_s=\pm1$ sublevels. By introducing an external magnetic field along the crystalline axis of the NV center, an additional Zeeman splitting between the $m_s=\pm1$ sublevels occurs. Thus we can isolate a two-level quantum system with sublevels $m_s=0$ and $-1$ (see Fig.~\ref{fig2}). The NV center can be described by the Hamiltonian~\cite{Neumann:2009}
%===============================================
\begin{equation}
H_{{\rm NV}} = D S_z^2+E(S_x^2-S_y^2)+g_e\mu\vec{B}\cdot\vec{S},
\label{NV}
\end{equation}
%===============================================
where $D$ is the ground-state zero-field splitting, $\vec{S}$ are the usual Pauli spin-1 operators, $E$ is the ground-state strain-induced splitting coefficient, $g_e=2$ is the NV Land\'e factor and $\mu=14$ MHzmT$^{-1}$ is the Bohr magneton. In this paper, we set $\hbar=1$. Furthermore, we consider the case where the strain-induced fine-structure splitting is negligible compared to the Zeeman splitting, i.e., $|E(S_x^2-S_y^2)|\ll|g_e\mu\vec{B}\cdot\vec{S}|$. Thus the second term in $H_{{\rm NV}}$ can be neglected here.

%===============================================
%===============================================

A flux qubit can have a superposition state of clockwise and counterclockwise persistent currents in the qubit loop with hundreds of nano amperes. By applying a static external magnetic field (with half a flux quantum perpendicular to the qubit loop), the flux qubit can be brought to the degeneracy point of the clockwise and counterclockwise persistent current states, where the qubit is less sensitive to the flux fluctuations. The flux qubit can be described by the Hamiltonian
%===============================================
\begin{equation}
H_{{\rm FL}} = \frac{1}{2}(\varepsilon\sigma_z+\lambda\sigma_x),
\label{FL}
\end{equation}
%===============================================
where $\vec{\sigma}$ denotes the Pauli operators of the flux qubit, $\lambda$ is the tunneling energy between the two wells of the qubit potential, and $\varepsilon=2I_p(\Phi-\Phi_0/2)$ is the energy bias of the flux qubit, with $I_p$ being its persistent current, $\Phi$ the applied magnetic flux, and $\Phi_0$ the magnetic-flux quantum. Obviously, parameters of the qubit, such as $I_p$ and $\lambda$, are determined by its fabrication, while the external magnetic flux $\Phi$ can be adjusted in the experiment. However, in a tunable four-junction flux qubit [see Fig.~\ref{fig1}(c)], $\lambda$ is a function of the external magnetic flux through the superconducting quantum interference device (SQUID),~\cite{Orlando:1999} and can also be tuned in experiments.

In our approach, we set the crystalline axis of the NV centers as the $z$-axis, and apply an external magnetic field $\vec{B}^{{\rm ext}}$, whose component parallel to the $z$-axis tunes the NV centers into near-resonance with the cavity mode, and whose component perpendicular to the qubit loop adjusts the superposition state of the clockwise and counterclockwise persistent current states of the flux qubit. These persistent currents produce an additional magnetic field $\vec{B}^{{\rm FQ}}$. The interaction between the NV center and the magnetic field produced by the flux qubit leads to a coupling between the two subsystems. The dynamics of these two coupled systems can be described by the Hamiltonian~\cite{Marcos:2010}
%===============================================
\begin{eqnarray}
H\!&\!=\!&\!\frac{1}{2}(\varepsilon\sigma_z+\lambda\sigma_x)\nonumber\\
\!&\!\!&\!+\sum_j^N\left[D\left(S_z^j\right)^2+g_e\mu B_z^{{\rm ext}}S_z^j+\sigma_z g_e\mu\vec{B}^{{\rm FQ}}\cdot\vec{S}^j\right],\nonumber\\
\end{eqnarray}
%===============================================
where $B_{z}^{{\rm ext}}$ is the parallel part of the external magnetic field, which adjusts the energy splitting of the NV center. When the $z$-axis is not perpendicular to the qubit loop, the frequencies of the NV center and the flux qubit can be adjusted by independently changing the components of the external magnetic field in different directions.~\cite{Marcos:2010} Here, we assume that the $z$-axis is parallel to the direction of the transmission-line resonator, as shown in Fig.~\ref{fig1}(a). The total Hamiltonian for the flux qubit and the two states with $m_s=0$ and $-1$ of the NV centers reads~\cite{Marcos:2010}
%===============================================
\begin{eqnarray}
H\!&\!=\!&\!\frac{1}{2}(\varepsilon\sigma_z+\lambda\sigma_x)\nonumber\\
\!&\!\!&\!+\sum_j^N\left[\frac{1}{2}\omega_{{\rm S}}\tau_z^j+ \frac{1}{\sqrt{2}}g_e\mu B^{{\rm FQ}}\sigma_z\left(\tau_+^j+\tau_-^j\right)\right],
\end{eqnarray}
%===============================================
where $\vec{\tau}$ denotes the Pauli operators of states with $m_s=0$ and $-1$ of the NV center, and $\omega_{{\rm S}}=D-g_e\mu B_z^{{\rm ext}}$ is the energy gap between these two states. The last term of this Hamiltonian describes the exchange of energy between NV centers and the flux qubit.

In order to enhance the coupling strength, a spin ensemble is employed rather than a single spin. The ground state of this ensemble is $|g\rangle=|0\cdots0\rangle$, while the excited state is $|e\rangle=\frac{1}{\sqrt{N}}\sum_j^N\tau_+^j|g\rangle$. We then define
%===============================================
\begin{equation}
s^{\dag}=\frac{1}{\sqrt{N}}\sum_j^N\tau_+^j
\end{equation}
%===============================================
to describe the collective excitation of the spin ensemble. In the conditions of large $N$ and low excitations, $s^{\dag}$ satisfies the bosonic commutation relations~\cite{Sun:2003}
%===============================================
\begin{equation}
[s,s^{\dag}]\approx1,
\end{equation}
%===============================================
and behaves as a bosonic operator, because only a few spins are excited. Therefore, the interaction between the flux qubit and the spin ensemble can be rewritten as
%===============================================
\begin{equation}
H_{{\rm QS}} = g_{{\rm QS}}\sigma_z(s^{\dag}+s),
\end{equation}
%===============================================
where $g_{{\rm QS}}=\sqrt{N}g_s$ is the coupling strength between the flux qubit and the spin ensemble, with 
%===============================================
\begin{equation}
g_s=\frac{1}{\sqrt{2}}g_e\mu B^{{\rm FQ}} 
\label{SIN}
\end{equation}
%===============================================
bing the coupling strength for a single NV center. As estimated in Ref.~\onlinecite{Marcos:2010}, the coupling strength can reach $g_{{\rm QS}}\sim 10$ MHz with $10^6$ NV centers, which is in the strong-coupling regime. Furthermore, by increasing the persistent current in the flux qubit, the coupling strength can be further enhanced.

\subsection{Flux qubit coupled to both a spin ensemble and a transmission-line resonator}

So far, a strong coupling between the flux qubit and the spin ensemble can be obtained.~\cite{Marcos:2010} Then, we integrate these two subsystems into a transmission-line resonator, as shown in Fig.~\ref{fig1}(a).

The transmission-line resonator has been realized in many experiments (see references in Ref.~\onlinecite{Schoelkopf:2008}). In this resonator, two ground planes are placed on the two sides of a central SC wire, and two gap capacitors at the two ends of the central wire play the role of ``mirrors'' in a conventional optical cavity. The distance between these two capacitors is an integer number of half-wavelengths. Such a structure forms a one-dimensional cavity with frequency $\sim$ 1--10 GHz when the entire setup is on the millimeter scale. The transmission-line resonator can be described by the Hamiltonian
%===============================================
\begin{equation}
H_{{\rm R}}=\omega_{{\rm R}}\left(a^{\dag}a+\frac{1}{2}\right),
\end{equation}
%===============================================
where $a~(a^{\dag})$ is the annihilation (creation) operator of the cavity, and $\omega_{{\rm R}}$ is the frequency of the cavity.

The flux qubit is fabricated at the antinode of the standing wave of the current on the transmission-line, where the strength of the magnetic field is maximum, so that at this place the flux qubit can strongly couple to the transmission-line resonator via the mutual inductance. The interaction between the flux qubit and the transmission-line resonator is described by the Hamiltonian~\cite{Bourassa:2009}
%===============================================
\begin{equation}
H_{{\rm QR}} = g_{{\rm QR}} \sigma_z(a^{\dag}+a),
\end{equation}
%===============================================
where $g_{{\rm QR}}=MI_pI_{r0}$ is the coupling strength between the flux qubit and the transmission-line resonator, with $I_{r0}=\sqrt{h\omega_{{\rm R}}/L_{{\rm R}}}$ being the zero-point current in the resonator and $L_{{\rm R}}$ the total inductance of the resonator.

In our proposed hybrid quantum circuit, the flux qubit and the central line of the resonator share a common segment for achieving strong coupling strength. In the case that the flux qubit separated from the central line of the resonator, the mutual coupling can only be enhanced by increasing the size of the qubit loop or reducing the distance between the qubit and the central line. However, a large area of the qubit loop would lead to a large susceptibility to the surrounding flux noise, while the close distance between the qubit and the central line induces an additional capacitive coupling between them. In contrast, the direct coupling via a shared segment does not have such problems and can reach a very strong coupling strength, even in the ultrastrong coupling regime when adding an additional Josephson junction on the central line of the resonator to increase the mutual inductance.~\cite{Niemczyk:2010}

Note that the NV center can also couple to the magnetic field in the transmission-line resonator.~\cite{Kubo:2010} However, compared with the magnetic field produced by the current in the qubit loop, this magnetic field is much weaker because of the reasons below: First, the current in the central line of the transmission-line resonator is usually smaller (about one order of magnitude or more smaller) than the current in the qubit loop. Second, a closed loop with a static current can produce a stronger magnetic field than the magnetic field at the same distance produced by the central line of the transmission-line resonator with a sinusoidal distributed current, when the maximum value of the current in the central line equals the static current of the qubit loop. Thus, the coupling strength between the NV center and the transmission-line resonator is much smaller (about two to three orders of magnitude smaller) than between the NV center and the flux qubit. Below we neglect this small interaction in our calculations.

%===============================================
\begin{figure}[tbp]
\includegraphics[width=2.75in]{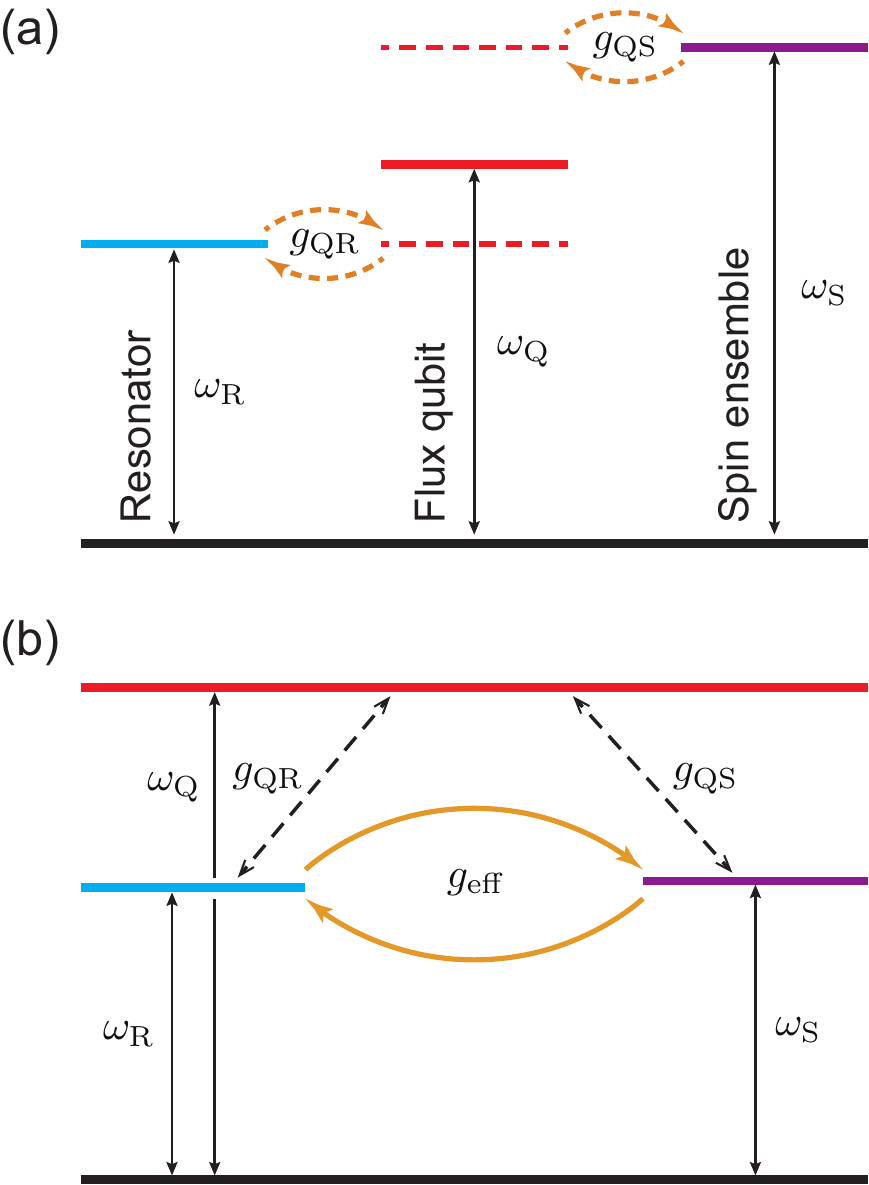}
\caption{(Color online) Schematic energy diagrams of the three subsystems in our proposed hybrid quantum circuit in two cases: (a) The tunable four-junction flux qubit acts as a data bus to exchange quantum information between the transmission-line resonator and the spin ensemble, whose frequencies $\omega_{\rm R}$ and $\omega_{\rm S}$ are fixed to far off-resonace. (b) The flux qubit is tuned to far off-resonance from the frequencies of the resonator and the spin ensemble for exchanging quantum information between these two subsystems via the virtual excitation (dashed black arrows) of the flux qubit. In these two diagrams, the blue horizontal segment refers to the transmission-line resonator, the red line denotes the flux qubit, and the purple horizontal line shows the ensemble of NV centers. The method (a) requires the accurate separate control of two couplings $g_{\rm QR}$ and $g_{\rm QS}$, while the approach in (b) requires the control of only one coupling $g_{\rm eff}$.}
\label{fig3}
\end{figure}
%===============================================

To the end, we adjust the flux qubit to the degeneracy point at $\varepsilon=0$ and express the Hamiltonian in the eigenvector basis of the flux qubit. Then, the total Hamiltonian of the proposed hybrid quantum circuit in Fig.~\ref{fig1}(a) can be written as
%===============================================
\begin{eqnarray}
H\!&\!=\!&\!\frac{1}{2}\omega_{{\rm Q}}\sigma_z+\omega_{{\rm R}}a^{\dag}a+\omega_{{\rm S}}s^{\dag}s\nonumber\\
\!&\!\!&\!+\,g_{{\rm QR}}(\sigma_++\sigma_-)(a^{\dag}+a) + g_{{\rm QS}}(\sigma_++\sigma_-)(s^{\dag}+s),\nonumber\\
\label{NJC}
\end{eqnarray}
%===============================================
where $\omega_{{\rm Q}}=\lambda$, and $\vec{\sigma}$ denotes the Pauli operators expressed in the eigenvector basis of the qubit. When the coupling strengths $g_{{\rm QR}}$ and $g_{{\rm QS}}$ are not very strong, i.e., the coupling strengths are much smaller than the frequencies of the cavity mode and the spin ensemble, the Hamiltonian can be reduced, in the rotating-wave approximation, into a Jaynes-Cummings form
%===============================================
\begin{eqnarray}
H\!&\!=\!&\!\frac{1}{2}\omega_{{\rm Q}}\sigma_z+\omega_{{\rm R}}a^{\dag}a+\omega_{{\rm S}}s^{\dag}s\nonumber\\
\!&\!\!&\!+\,g_{{\rm QR}}(\sigma_+a+\sigma_-a^{\dag}) + g_{{\rm QS}}(\sigma_+s+\sigma_-s^{\dag}).
\label{JC}
\end{eqnarray}
%===============================================
Note that the resonant case, i.e., $\omega_{{\rm Q}}=\omega_{{\rm R}}=\omega_{{\rm S}}$, was theoretically discussed in Ref.~\onlinecite{Twamley:2010} using a similar hybrid circuit, where only a single NV center was employed. If a tunable four-junction flux qubit is used as in our approach [see Fig.~\ref{fig1}(c)], we can transfer the information between the spin ensemble and the transmission-line resonator following the steps below [see Fig.~\ref{fig3}(a)]. First, we fix the frequencies of the transmission-line resonator and the spin ensemble to far off-resonance. Then, by changing the magnetic flux through the SQUID, we can adjust the frequency of the flux qubit successively into resonance with the resonator and the spin ensemble to achieve the quantum-information transmission from the resonator to the flux qubit and then to the spin ensemble, and vice versa. Here the flux qubit acts as a data bus. However, for a high-fidelity quantum-information transmission with the above protocol, it requires very accurate time-dependent controls for the coupling between the flux qubit and the transmission-line resonator (spin ensemble).

To avoid using these very accurate time-dependent controls of the two couplings, we focus on the case when the flux qubit is tuned to have a large qubit energy, so as to be far off-resonance from the frequencies of the transmission-line resonator and the spin ensemble [see Fig.~\ref{fig3}(b)]. Here the resonator and the spin ensemble are assumed to be near resonance to each other. Thus, an effective interaction between the resonator and the spin ensemble, with coupling strength $g_{\rm eff}$, can be achieved by adiabatically eliminating the degrees of freedom of the flux qubit. Choosing appropriate parameters of the circuit, this effective coupling strength can be much larger than the direct-coupling strength between the transmission-line resonator and the spin ensemble. For details, see the two sections below. Importantly, the information-transmission protocol based on this effective interaction does not require very accurate time-dependent controls of the two coupling strengths $g_{{\rm QR}}$ and $g_{{\rm QS}}$. One could transfer quantum information between the resonator and the spin ensemble only by controlling the effective coupling strength $g_{\rm eff}$.

\section{Strong-coupling regime}\label{sec:3}

We now consider the case where the flux qubit strongly couples to both a transmission-line resonator and a spin ensemble, i.e., $\kappa, \gamma\ll g_{{\rm QR}},g_{{\rm QS}}\ll \omega_{{\rm R}},\omega_{{\rm S}}$. Moreover, the frequency $\omega_{{\rm Q}}$ of the flux qubit is fixed to be much larger than the frequency $\omega_{{\rm R}}$ ($\omega_{{\rm S}}$) of the resonator (spins) and satisfies $\Delta_{{\rm R}},\Delta_{{\rm S}}\gg g_{{\rm QR}},g_{{\rm QS}}$, where $\Delta_{{\rm R(S)}}=\omega_{{\rm Q}}-\omega_{{\rm R(S)}}$. This corresponds to the large detuning regime and allows us to apply a Fr\"ohlich-Nakajima transformation~\cite{Frohlich:1950,Nakajima:1953} to deduce an effective coupling between the spin ensemble and the transmission-line resonator. Here we further assume that the frequencies of the transmission-line resonator and spins are both slightly off-resonance to each other.

We now rewrite the total Hamiltonian (\ref{JC}) as $H=H_0+H_I$ in terms of the free part
%===============================================
\begin{equation}
H_0=\frac{1}{2}\omega_{{\rm Q}}\sigma_z+\omega_{{\rm R}}a^{\dag}a+\omega_{{\rm S}}s^{\dag}s\;,
\label{FR}
\end{equation}
%===============================================
and the interaction part
%===============================================
\begin{equation}
H_I=g_{{\rm QR}}(\sigma_+a+\sigma_-a^{\dag}) + g_{{\rm QS}}(\sigma_+s+\sigma_-s^{\dag}).
\end{equation}
%===============================================
To use the Fr\"ohlich-Nakajima transformation,~\cite{Frohlich:1950,Nakajima:1953} we should find out a unitary transformation $U=\exp(-V)$, such that $V$ is an anti-Hermitian operator $V=-V^{\dag}$ and satisfies
%===============================================
\begin{equation}
H_I+[H_0,V]=0.
\end{equation}
%===============================================
We apply this unitary transformation to $H$ and obtain, up to second order, an effective Hamiltonian
%===============================================
\begin{equation}
H_{{\rm eff}}=UHU^{\dag}=H_0+\frac{1}{2}[H_I,V]+O(g^3).
\end{equation}
%===============================================
In the present case, the anti-Hermitian operator $V$ for the Fr\"ohlich-Nakajima transformation adopts the following form:
%===============================================
\begin{equation}
V=\xi_{{\rm R}}(\sigma_-a^{\dag}-\sigma_+a)+\xi_{{\rm S}}(\sigma_-s^{\dag}-\sigma_+s)\;,
\end{equation}Ê
%===============================================
where $\xi_{{\rm R}}=g_{{\rm QR}}/\Delta_{{\rm R}}$, and $\xi_{{\rm S}}=g_{{\rm QS}}/\Delta_{{\rm S}}$.

Because the coefficients $\xi_{{\rm R}}$ and $\xi_{{\rm S}}$ are small in the large detuning regime, the high-order terms of the Fr\"ohlich-Nakajima transformation can be dropped out and only the second-order term $[H_I,S]$ should be considered. Furthermore, we assume that the flux qubit is initially in the ground state. The interaction between the resonator and spins is induced by virtual excitation of the flux qubit, without real energy exchanges between the flux qubit and the two subsystems. Thus, we can eliminate the degrees of freedom of the flux qubit, and obtain the effective Hamiltonian as
%===============================================
\begin{equation}
H_{{\rm eff}}=\omega_{{\rm R}}^{\prime}a^{\dag}a+\omega_{{\rm S}}^{\prime}s^{\dag}s+g_{{\rm eff}}(a^{\dag}s+as^{\dag})\;,
\end{equation}
%===============================================
where
%===============================================
\begin{eqnarray}
\omega_{{\rm R}}^{\prime}\!&\!=\!&\!\omega_{{\rm R}}-\frac{g_{{\rm
QR}}^2}{\Delta_{{\rm R}}}\;,~~~ \omega_{{\rm S}}^{\prime}=\omega_{{\rm
S}}-\frac{g_{{\rm QS}}^2}{\Delta_{{\rm S}}}\;, \\
g_{{\rm eff}}\!&\!=\!&\!-\frac{1}{2}\left(\frac{1}{\Delta_{{\rm
R}}}+\frac{1}{\Delta_{{\rm S}}}\right)\,g_{{\rm QR}}\,g_{{\rm QS}}.
\label{eff-1}
\end{eqnarray}
%===============================================

According to the experimental data in Ref.~\onlinecite{Zhu:2011}, we can choose the coupling strength between an NV center and a flux qubit as $\sim 12$ kHz. When the number of spins in the ensemble is $\sim7\times10^7$, the coupling strength $g_{\rm QS}$ between the flux qubit and the spin ensemble is approximately $100$~MHz. Here we assume that the coupling strength $g_{\rm QR}$ between the flux qubit and the transmission-line resonator is also approximately equal to $100$~MHz, and the detuning between the flux qubit and the resonator (spins) is $\sim 1$~GHz. From Eq.~(\ref{eff-1}), the effective coupling strength $g_{\rm eff}$ is estimated to be $g_{{\rm eff}}\sim 10$ MHz, which is comparable to the direct-coupling strength between $10^{12}$ NV centers and the transmission-line resonator (a recent experiment is in Ref.~\onlinecite{Kubo:2010}). The low decay rates from the cavity ($\kappa\sim$ 1 MHz), the flux qubit ($\gamma_{\rm Q}\sim$ 1MHz), and the spin ensemble (1 $<\gamma_{\rm S}<$ 10 MHz) have been implemented in recent experiments.~\cite{Kubo:2010} Thus, this effective coupling is in the strong coupling regime.  This strong coupling can be used to transfer quantum information between the spin ensemble and photon states, which can act as flying qubits for quantum communication with other systems, such as the SC qubit, in future hybrid quantum circuits. Since the flux qubit is always in its ground state, its decoherence would not affect the quantum state transfer.

%===============================================
\begin{figure}
\includegraphics[width=3.35in,bb=39 14 423 311]{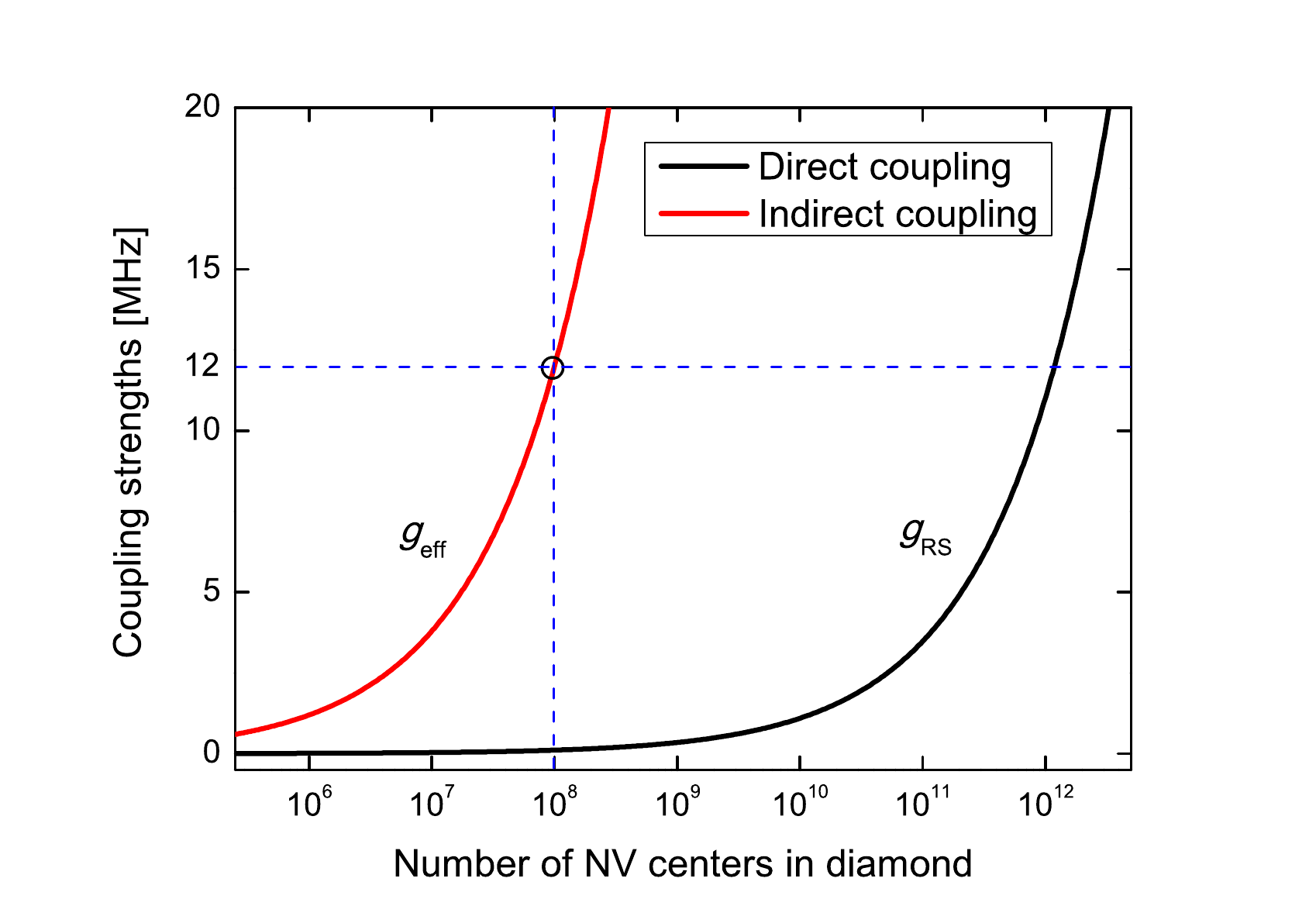}
\caption{(Color online) Two kinds of coupling strengths between an ensemble of NV centers and the transmission-line resonator, as a function of the number of NV centers in diamond. The red curve on the left shows the effective coupling $g_{\rm eff}$ between the spin ensemble and the resonator. The data used in this curve for a single NV center are from Ref.~\onlinecite{Zhu:2011}. The black curve on the right describes the direct coupling $g_{\rm RS}$ between the spin ensemble and the resonator. The data used in this curve for a single NV center is from Ref.~\onlinecite{Kubo:2010}. The blue dashed lines denote the parameter value (number of NV centers $=10^8$) for observing strong coupling strength (12 MHz) in our approach. Note that for $N\sim10^9$, $g_{\rm eff}\sim 10^2$ MHz.}
\label{fig4}
\end{figure}
%===============================================

Figure~\ref{fig4} shows the effective-coupling strength $g_{{\rm eff}}$ (with single-NV-center coupling data from Ref.~\onlinecite{Zhu:2011}) in our approach and the direct-coupling strength $g_{{\rm RS}}$ (with single-NV-center coupling data from Ref.~\onlinecite{Kubo:2010}) versus the number of NV centers in diamond. Fig.~\ref{fig4} clearly shows that by using an ensemble with the same number of NV centers (spins), our approach can implement much larger coupling strength, as compared to the direct-coupling approach. Physically, this very enhanced coupling strength is induced by the two strong couplings between the flux qubit and the other subsystems (the transmission-line resonator and the ensemble of NV centers).

Due to the presence of interference effects caused by the inhomogeneous broadening, the fidelity (which describes the correspondence of the readout signal with the original signal) was low and the coherence times were not very long when the density of NV centers in diamond is high.~\cite{Acosta:2009,Stanwix:2010} There are two main inhomogeneous broadenings leading to dephasing of the spin ensemble of NV centers. One is the inhomogeneous broadening due to dipolar hyperfine coupling of nearby $^{13}$C nuclear spins, which might be reduced by polarizing the nuclear spins. The other one is the dipolar broadening due to the residual paramagnetic nitrogen atoms. Because the residual paramagnetic nitrogen atoms have a density several times higher than the NV centers in a given sample, this dipole broadening is the dominant dephasing mechanism in the high-nitrogen-concentration diamond crystals.~\cite{Acosta:2009,Kurucz:2011,Diniz:2011} In general, decreasing the density of NV centers during sample preparation is accompanied by a decreased density of residual nitrogen paramagnetic impurities, which can reduce the dephasing from the second inhomogeneous broadening. Therefore, a diamond sample with a low-density of NV centers could improve the coherence performance of the spin ensemble. 

To achieve a strong coupling strength, our approach requires far less NV centers. Because the effective size of the spin ensemble used in our proposal is much smaller than that utilized in the direct-coupling approach, the spin densities of these two different approaches in recent experiments are comparable.~\cite{Zhu:2011,Kubo:2010} However, we can still enhance the coupling strength of a single NV center by either using a flux qubit with a larger persistent current or changing the shape of the flux-qubit loop, and then further reduce the number of NV centers needed in our circuit.

Now we estimate the coupling strength in our proposed circuit by considering a realistic NV-center sample. As in Ref.~\onlinecite{Zhu:2011}, we choose a rectangular loop for the flux qubit. According to the Biot-Savart law, the magnetic field in the center of the rectangular loop generated by the persistent current of a flux qubit can be written as
%===============================================
\begin{equation}
B^{\rm FQ} = \alpha\frac{\mu_0I_p}{4\pi\sqrt{A}},
\end{equation}
%===============================================
where $\alpha=8\sqrt{\beta+1/\beta}$, with $\beta$ being the length-width ratio of the rectangular loop, $A$ is the area of the loop, $\mu_0=4\pi\times10^{-7}$N$\cdot$A$^{-2}$ is the vacuum permeability, and $I_p$ denotes the persistent current of the flux qubit. From Eq.~(\ref{SIN}), it follows that the coupling strength $g_s$ is
%===============================================
\begin{equation}
g_s=\alpha \frac{g_e\mu\mu_0I_p}{4\pi\sqrt{2A}}.
\end{equation}
%===============================================
Thus, the coupling strength between the flux qubit and the spin ensemble can be estimated as
%===============================================
\begin{equation}
g_{\rm QS}=\sqrt{DV}g_s\approx\sqrt{D\,d}\,\alpha g_e\mu\mu_0I_p,
\label{GQS}
\end{equation}
%===============================================
where $D$ is the density of NV centers within the rectangular loop, and $V=A_{\rm S}d$ is the volume of NV centers that effectively couple to the flux qubit, with $A_{\rm S}\approx A$ and $d$ being the thickness of the NV-center sample. We consider our circuit with experimentally accessible parameters: $D\sim3\times10^6~\mu$m$^{-3}$ (see Ref.~\onlinecite{Acosta:2009}), $I_p\sim 900$ nA (see Ref.~\onlinecite{Paauw:2009}), $\beta\sim50$ (see Ref.~\onlinecite{Zhu:2011}), and $d\sim5~\mu$m. The coupling strength between the flux qubit and the NV centers is estimated as $g_{\rm QS}\sim 350$ MHz. Therefore, according to Eq.~(\ref{eff-1}), the effective coupling strength $g_{\rm eff}$ between the spin ensemble and the resonator can be $\sim120$ MHz, when $g_{\rm QR}\approx350$ MHz and $\Delta_{\rm {R(S)}}\sim3g_{\rm QR(QS)}$, which is much larger than the direct-coupling strength $\sim10$ MHz (see Ref.~\onlinecite{Xiang:2013}). If the effective coupling strength is chosen as $\sim 10$ MHz, the NV-center density of the sample is reduced to $D\sim2\times10^4~\mu$m$^{-3}$, which is much lower than the NV-center density $D\sim3\times10^6~\mu$m$^{-3}$ for achieving the same value of the direct-coupling strength. As shown in Refs.~\onlinecite{Acosta:2009,Stanwix:2010,BarGill:2012}, a lower density of NV centers in the sample can improve the quantum coherence of the spin ensemble. In fact, the magnetic field close to the edge of the flux-qubit loop is much larger than that in the center of the loop, so the real value of $g_{\rm eff}$ should be larger than the value estimated above. 

The coherence performance of the spin ensemble is affected not only by the width of the inhomogeneous broadening but also by its shape.~\cite{Kurucz:2011,Diniz:2011} By choosing an appropriate type of distribution of the inhomogeneous broadening of spins, such as a Gaussian distribution, the decoherence would be dominated by the spins' homogeneous broadening. This effect, known as cavity protection, could provide longer coherence times in our proposed circuit.

\section{Ultrastrong-coupling regime}\label{sec:4}

Recently, the coupling strength between the flux qubit and the transmission-line resonator experimentally reached the ultrastrong-coupling regime~\cite{Niemczyk:2010} [$g_{\rm QR}/\omega_{\rm R}\approx 0.1$]. When the number of spins is larger than $\sim10^{9}$, the coupling strength between the flux qubit and the spin ensemble could also be in the ultrastrong-coupling regime. Note that, in this case, the density of NV centers in the sample is $>10^7~\mu$m$^{-3}$, which is still larger than the density of NV centers achieved in recent experiments.~\cite{Acosta:2009}

In such an ultrastrong-coupling regime, the Hamiltonian of our proposed hybrid quantum circuit in Fig.~\ref{fig1}(a) cannot be reduced into the simple Jaynes-Cummings form, because the counter-rotating terms cannot be neglected, and the higher-order terms in the Fr\"ohlich-Nakajima transformation cannot be dropped in some cases. Here, we use again the Fr\"ohlich-Nakajima transformation to analyze the dynamics of our proposed circuit when the large detuning condition is satisfied.

We write the total Hamiltonian in terms of the free part $H_0$ and the interaction part $H_I$. The free part $H_0$ has the same form as Eq.~(\ref{FR}), but the interaction part should be written, without the rotating-wave approximation, as
%===============================================
\begin{equation}
H_I=g_{{\rm QR}}(\sigma_++\sigma_-)(a^{\dag}+a) + g_{{\rm QS}}(\sigma_++\sigma_-)(s^{\dag}+s).
\end{equation}
%===============================================
In this case, the anti-Hermitian operator $V$ for the Fr\"ohlich-Nakajima transformation adopts the form:
%===============================================
\begin{eqnarray}
V\!&\!=\!&\!\xi_{{\rm R}}(\sigma_-a^{\dag}-\sigma_+a)+\zeta_{{\rm R}}(\sigma_-a-\sigma_+a^{\dag})\nonumber\\
\!&\!\!&\!+\xi_{{\rm S}}(\sigma_-s^{\dag}-\sigma_+s)+\zeta_{{\rm S}}(\sigma_-s-\sigma_+s^{\dag})\;,
\end{eqnarray}
%===============================================
where $\zeta_{{\rm R}}=g_{{\rm QR}}/\eta_{{\rm R}}$, $\zeta_{{\rm S}}=g_{{\rm QS}}/\eta_{{\rm S}}$, and $\eta_{{\rm R(S)}}=\omega_{{\rm Q}} +\omega_{{\rm R(S)}}$.

After eliminating the degrees of freedom of the flux qubit, we can obtain an effective Hamiltonian as follows:
%===============================================
\begin{eqnarray}
H_{{\rm eff}}\!&\!=\!&\!\omega_{{\rm R}}^{\prime}a^{\dag}a+\omega_{{\rm S}}^{\prime}s^{\dag}s+g_{{\rm eff}}(a^{\dag}+a)(s^{\dag}+s)\nonumber\\
\!&\!\!&\!-\frac{1}{2}\alpha_{{\rm R}}\,g_{{\rm QR}}^2(a^{\dag}a^{\dag}+aa)-\frac{1}{2}\alpha_{{\rm S}}\,g_{{\rm QS}}^2(s^{\dag}s^{\dag}+ss)\;,\nonumber\\
\!&\!\!&\!+\,\mathcal{O}
\label{ultra-Heff}
\end{eqnarray}
%===============================================
where 
%===============================================
\begin{equation}
\omega_{{\rm R}}^{\prime}=\omega_{{\rm R}}-\alpha_{{\rm R}}\,g_{{\rm QR}}^2\;, \quad\omega_{{\rm S}}^{\prime}=\omega_{{\rm S}}-\alpha_{{\rm S}}\,g_{{\rm QS}}^2\;,
\end{equation}
%===============================================
\begin{equation}
g_{{\rm eff}}=-(\alpha_{{\rm R}}+\alpha_{{\rm S}})\,g_{{\rm QR}}\,g_{{\rm QS}}/2\;,
\end{equation}
%===============================================
\begin{equation}
\alpha_{{\rm R(S)}}=\frac{1}{\Delta_{{\rm R(S)}}}+\frac{1}{\eta_{{\rm R(S)}}},
\end{equation}
%===============================================
and $\mathcal{O}$ represents the higher-order terms that can be neglected when $g/\Delta$ is small. Owing to the larger coupling strength $g_{{\rm QR}}$ and $g_{{\rm QS}}$, the effective coupling strength $g_{{\rm eff}}$ can be larger than that in the strong-coupling regime derived in Sec.~\ref{sec:3}, but additional terms appear in the second line of Eq.~(\ref{ultra-Heff}), which can produce nonlinear effects in the system.

%===============================================
\begin{figure}[tbp]
\includegraphics[width=3.3in]{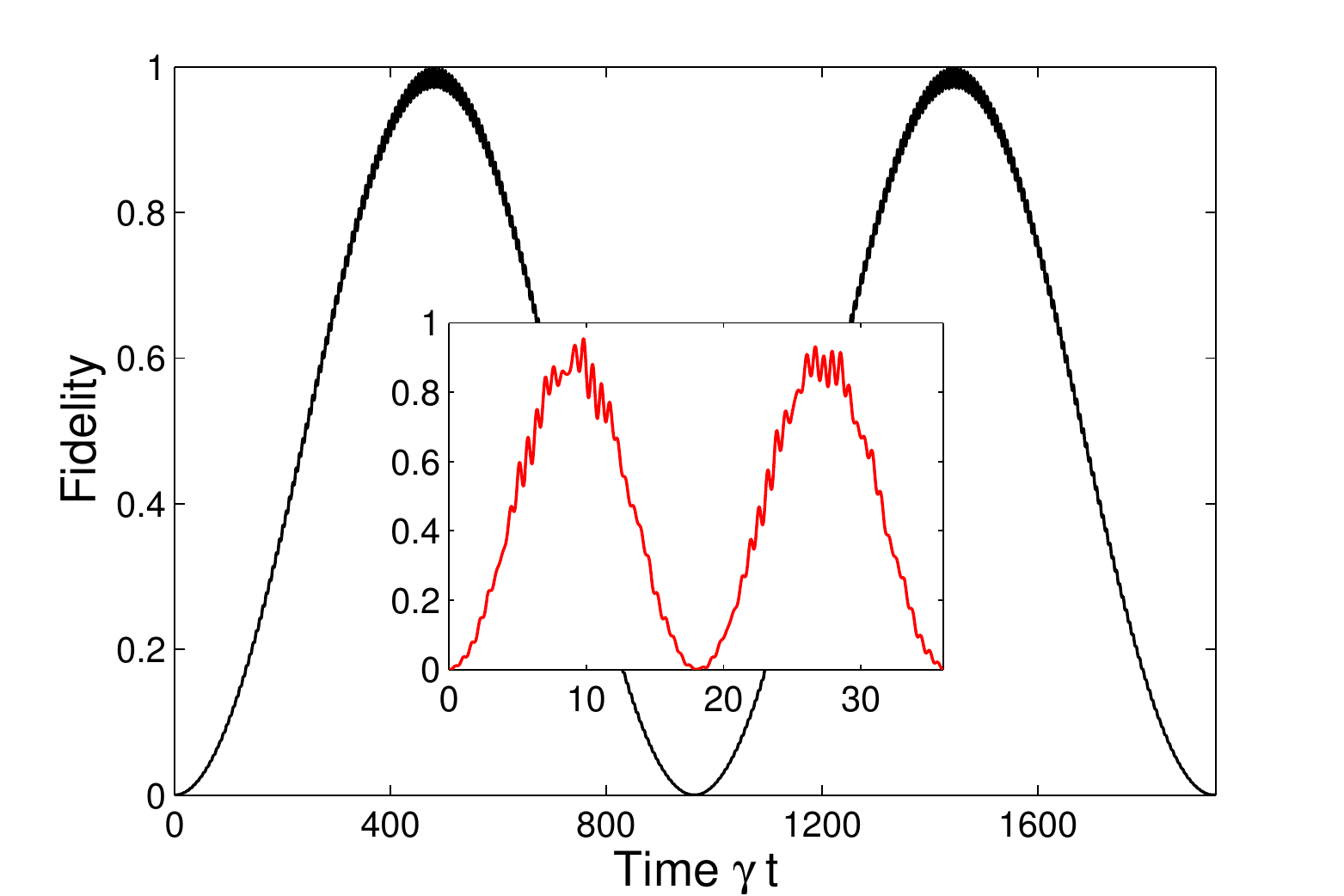}
\caption{(Color online) The fidelity of quantum state transfer versus the dimensionless time $\gamma t$. The red and black curves correspond to the coupling strength in the ultrastrong-coupling regime ($g_{\rm QR}=g_{\rm QS}=\omega_{\rm R}=\omega_{\rm S}$ and $\omega_{\rm Q}=9\,\omega_{\rm R}$) and the strong-coupling regime ($g_{\rm QR}=g_{\rm QS}=0.05\,\omega_{\rm R},~\omega_{\rm S}=\omega_{\rm R}$, and $\omega_{\rm Q}=2\,\omega_{\rm R}$), respectively.}
\label{fig5}
\end{figure}
%===============================================

\begin{figure}[tbp]
\includegraphics[width=3.1in]{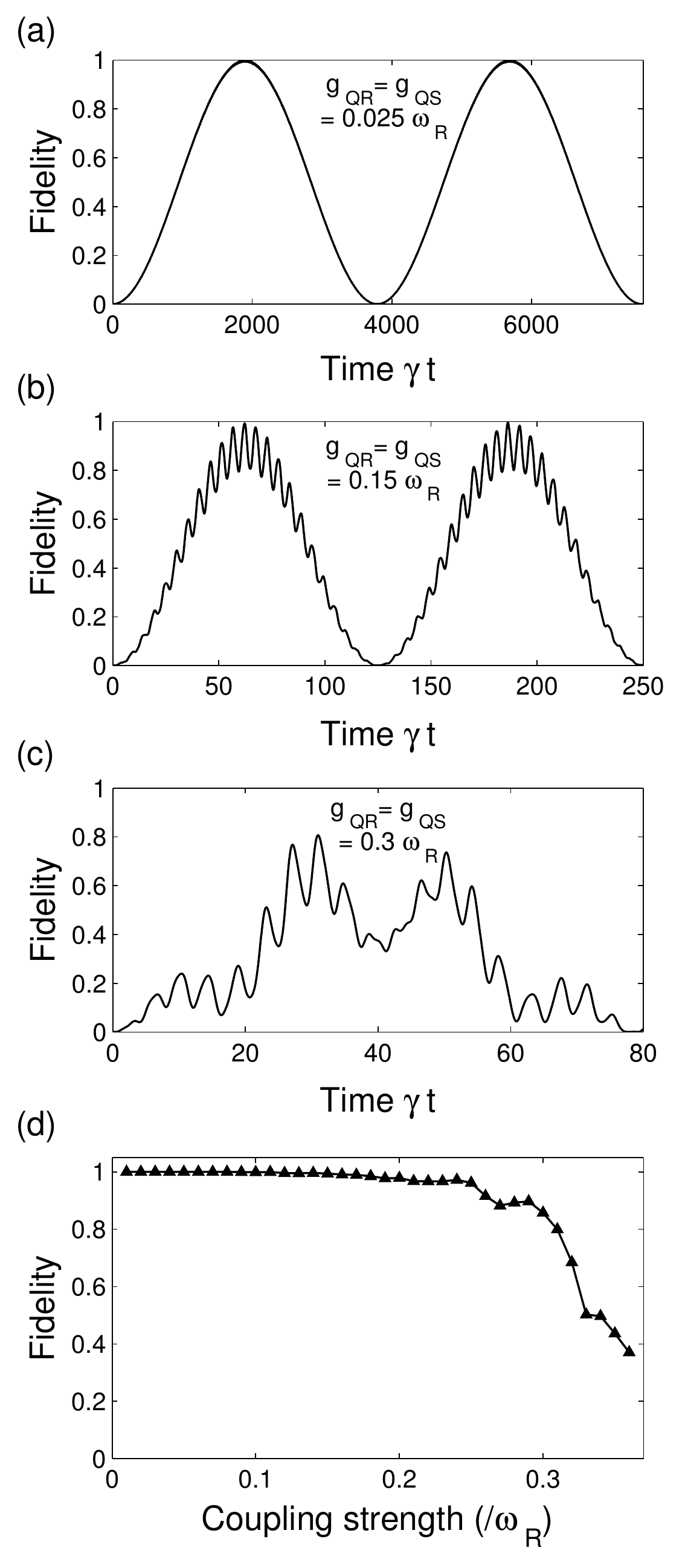}
\caption{The fidelity of quantum state transfer vs the dimensionless time $\gamma t$ with the coupling strength (a) $g_{\rm QR}=g_{\rm QS}=0.025\,\omega_{\rm R}$, (b) $g_{\rm QR}=g_{\rm QS}=0.15\,\omega_{\rm R}$, and (c) $g_{\rm QR}=g_{\rm QS}=0.3\,\omega_{\rm R}$. (d) The fidelity of quantum state transfer versus the coupling strength ($/\omega_{\rm R}$). In these four figures, the frequencies of the transmission-line resonator and the spin ensemble are resonant, $\omega_{\rm R}=\omega_{\rm S}$, and the frequency of the flux qubit is $\omega_{\rm Q}=\,2\omega_{\rm R}$.}
\label{fig6}
\end{figure}

The first term in the second line involves squeezed photon states in the resonator. We can apply a unitary transformation on the Hamiltonian~\cite{Yuen:1976}
%===============================================
\begin{equation}
U=\exp\left[\frac{1}{2}r^*a^2-\frac{1}{2}r(a^{\dag})^2\right],
\end{equation}
%===============================================
and obtain that
%===============================================
\begin{eqnarray}
H_{{\rm eff}}\!&\!=\!&\!\left[\omega_{{\rm R}}^{\prime}(\sinh^2r+\cosh^2r)+2\alpha_{{\rm R}}g_{{\rm QR}}^2\sinh r\cosh r\right]a^\dag a\nonumber\\
\!&\!\!&\!+\omega_{{\rm S}}^{\prime}s^{\dag}s-\frac{1}{2}\alpha_{{\rm S}}g_{{\rm QS}}^2(s^{\dag}s^{\dag}+ss)\nonumber\\
\!&\!\!&\!+g_{{\rm eff}}(\cosh r-\sinh r)(a^{\dag}+a)(s^{\dag}+s)\;,
\end{eqnarray}
%===============================================
where $r$ satisfies the equation $$\sinh^2r=\frac{2}{\sqrt{\beta^2-4}\,(\sqrt{\beta^2-4}+\beta)},$$ with $\beta=2\omega_{{\rm R}}/\alpha_{{\rm R}}g_{{\rm QR}}^2-2$. Thus, the energy exchange in this case is between the squeezed photon states in the resonator and the collective excitations of the spin ensemble. The second term in the second line of Eq.~(\ref{ultra-Heff}) breaks the low-excitation condition for the bosonic operator of the spin ensemble. These lead to an obvious reduction of the fidelity of the quantum state transfer, see Fig.~\ref{fig5}. Here the fidelity is defined as $|\langle\psi_{\rm T}|\psi(t)\rangle|^2$, where $|\psi_{\rm T}\rangle$ is the target state of quantum transfer. Note that, in order to clearly show the effect on the fidelity from the change of the coupling strength, we neglect the decoherence that comes from both the inhomogeneous broadening of the spin ensemble and the photon leaking of the cavity. This decoherence is also neglected in Fig.~\ref{fig6}.

%===============================================
%===============================================

When the coupling strengths, $g_{\rm QR}$ and $g_{\rm QS}$, reach the ultrastrong-coupling regime, in order to satisfy the large-detuning condition the energy gap of the eigenstates of the flux qubit should be large enough, compared with the frequency of the spin ensemble (the resonator). A large-gap flux qubit can be achieved by using Josephson junctions with larger Josephson energy $E_J$ or by adjusting the qubit to move away the degeneracy point of the persistent current states. However, this induces large flux noise and results in decoherence of the flux qubit. In general, the frequency of the flux qubit is several times that of the NV center. Here we numerically calculated the fidelity of the quantum state transfer by changing the coupling strength when $\omega_{\rm Q}=2\omega_{\rm S}$, as shown in Fig.~\ref{fig6}. In this case, while the coupling strength is increasing, the fidelity is decreasing. After the large detuning condition is broken, the fidelity rapidly reduces to a low level, because the terms $\mathcal{O}$ in Eq.~(\ref{ultra-Heff}) cannot be neglected anymore. Note that our numerical calculations neglected the effect from the decays of the resonator and spins; otherwise the fidelity should be much lower when the coupling strength is in the weak-coupling regime.

To keep the low-excitation condition satisfied, we consider another case where the flux qubit ultrastrongly couples to the resonator ($g_{\rm QR}\sim 0.1 \omega_{\rm R}$), but strongly couples to the spin ensemble. In such a case, the interaction part of the Hamiltonian becomes
%===============================================
\begin{equation}
H_I=g_{{\rm QR}}(\sigma_++\sigma_-)(a^{\dag}+a) + g_{{QS}}(\sigma_+s+\sigma_-s^{\dag}),
\end{equation}
%===============================================
and the corresponding anti-Hermitian operator $V$ in the Fr\"ohlich-Nakajima transformation has the form
%===============================================
\begin{eqnarray}
V\!&\!=\!&\!\xi_{{\rm R}}(\sigma_-a^{\dag}-\sigma_+a)+\zeta_{{\rm R}}(\sigma_-a-\sigma_+a^{\dag})\nonumber\\
\!&\!\!&\!+\xi_{{\rm S}}(\sigma_-s^{\dag}-\sigma_+s).
\end{eqnarray}Ê
%===============================================

In applying the Fr\"ohlich-Nakajima transformation, the coefficients of some terms in the third order can be comparable to those of the terms in the second order. With these terms retained, the effective Hamiltonian becomes
%===============================================
\begin{eqnarray}
H_{{\rm eff}}\!&\!=\!&\!\omega_{{\rm R}}^{\prime}a^{\dag}a+\omega_{{\rm S}}^{\prime}s^{\dag}s\nonumber\\
\!&\!\!&\!-\frac{g_{{\rm QR}}g_{{\rm QS}}}{2\Delta_{{\rm S}}}(a^{\dag}+a)(s^{\dag}+s)\nonumber\\
\!&\!\!&\!-\frac{g_{{\rm QR}}g_{{\rm QS}}}{2\Delta_{{\rm R}}}(a^{\dag}s+as^{\dag})-\frac{g_{{\rm QR}}g_{{\rm QS}}}{2\eta_{{\rm R}}}(a^{\dag}s^{\dag}+as)\nonumber\\
\!&\!\!&\!-\alpha_{{\rm R}}\frac{2g_{{\rm QR}}^3}{3\eta_{{\rm R}}}(\sigma_+a+\sigma_-a^{\dag}+2\sigma_+a^{\dag}+2\sigma_-a)\nonumber\\
\!&\!\!&\!-\alpha_{{\rm R}}\frac{2g_{{\rm QR}}^3}{3\Delta_{{\rm R}}}(\sigma_+a^{\dag}+\sigma_-a+2\sigma_+a+2\sigma_-a^{\dag})\nonumber\\
\!&\!\!&\!-\alpha_{{\rm R}}\frac{2g_{{\rm QR}}^3}{3\eta_{{\rm R}}}(\sigma_+a^{\dag}a^{\dag}a^{\dag}+\sigma_+a^{\dag}aa+2\sigma_+a^{\dag}a^{\dag}a\nonumber\\
\!&\!\!&\!+\sigma_-aaa+\sigma_-a^{\dag}a^{\dag}a+2\sigma_-a^{\dag}aa)\nonumber\\
\!&\!\!&\!-\alpha_{{\rm R}}\frac{2g_{{\rm QR}}^3}{3\Delta_{{\rm R}}}(\sigma_-a^{\dag}a^{\dag}a^{\dag}+\sigma_-a^{\dag}aa+2\sigma_-a^{\dag}a^{\dag}a\nonumber\\
\!&\!\!&\!+\sigma_+aaa+\sigma_+a^{\dag}a^{\dag}a+2\sigma_+a^{\dag}aa) + \mathcal{O},
\end{eqnarray}
%===============================================
where $\mathcal{O}$ represents the higher-order terms that can be neglected. In this Hamiltonian, the first line gives the effective energies of both photons and spins, the second and third lines involve the energy exchange between the resonator and the spin ensemble, the next two lines describe the energy exchange between the flux qubit and the resonator, and the last four lines describe the energy exchange between the flux qubit and the resonator involving three-photon processes. 

As a result of this, the photon state will not be confined in the subspace $\{|0\rangle,|1\rangle\}$. Also, the flux qubit will not remain in the ground state as we initially assumed, because the transimission-line resonator can now exchange energy not only with the spin ensemble, but also with the flux qubit. Thus, the total system will exhibit very rich quantum-dynamical behavior. However, the fidelity of the quantum state transfer would obviously be very low in this case; thus this circuit would not be suitable to exchange the state between the resonator and the spin ensemble in this parameter regime.

\section{Discussion and conclusion}\label{sec:5}

Here we emphasize that the strong coupling $g_{{\rm QR}}$ between a flux qubit and a transmission-line resonator~\cite{Niemczyk:2010} and the strong coupling $g_{{\rm QS}}$ between a flux qubit and an ensemble of NV centers (spins)~\cite{Zhu:2011} have both been achieved in experiments. Therefore, it becomes feasible to construct our proposed hybrid quantum circuit, to realize a strong coupling between the spin ensemble and the resonator. This strong coupling can be used to transfer quantum information between the spin ensemble (as a quantum memory) and photon states (as flying qubits). Moreover, due to the much smaller number of NV centers used in our approach, it is expected that a low-density sample of NV centers could be adopted for better coherent performance. In addition, the ultrastrong-coupling regime has recently become a very attractive topic, and the corresponding case in our proposed circuit is also discussed. Because the Hamiltonian in this regime becomes more complex, rich quantum-dynamical phenomena are expected and these will be explored in the future. 

In conclusion, we have proposed an approach to achieve a very strong effective coupling between a spin ensemble and a transmission-line resonator via a flux qubit. Our approach provides an experimentally realizable hybrid circuit for exchanging quantum information between a SC resonator and a low-density spin ensemble with long-coherence time. Also, our proposed circuit can be fabricated on a chip, facilitating its future scalability, which is crucial for future quantum technologies.

\acknowledgements The authors would like to thank Yuimaru Kubo for useful comments and suggestions. ZLX and JQY were partly supported by the National Basic Research Program of China Grant No.2009CB929302, NSFC Grant No.91121015, and MOE Grant No.B06011. ZLX was also partly supported by the RIKEN IPA program. FN was supported in part by the ARO, JSPS-RFBR Grant 12-02-92100, Grant-in-Aid for Scientific Research (S), MEXT Kakenhi on Quantum Cybernetics, and the JSPS through the FIRST program. XYL was supported by Japanese Society for the Promotion of Science (JSPS) Foreign Postdoctoral Fellowship No. P12204 and the NSF of China under grant number 11005057. TFL was partly supported by NSFC grant No. 61106121 and No. 61174084.

\end{document}